\documentclass[conference,letterpaper]{IEEEtran}
\IEEEoverridecommandlockouts
\usepackage{cite}
\usepackage{amsmath,amssymb}
\usepackage{newtxtext,newtxmath}
\usepackage{graphicx}
\usepackage{booktabs}
\usepackage{url}
\usepackage[hidelinks]{hyperref}
\graphicspath{{figures/}}
\AtBeginDocument{%
  \setlength{\abovedisplayskip}{4pt plus 1pt minus 1pt}%
  \setlength{\belowdisplayskip}{4pt plus 1pt minus 1pt}%
  \setlength{\abovedisplayshortskip}{2pt plus 1pt minus 1pt}%
  \setlength{\belowdisplayshortskip}{2pt plus 1pt minus 1pt}%
}

\IEEEaftertitletext{\vspace{-2\baselineskip}} 
\begin{document}

\title{Adaptive Grid-Mode-Aware Hybrid Energy Storage\\
Control for AI Data Center Oscillation Mitigation}
\author{\IEEEauthorblockN{Quang Manh Hoang\textsuperscript{1}, Sina Mohammadi\textsuperscript{1},
Phi Long Nguyen\textsuperscript{2,3}, Wencong Su\textsuperscript{1}, Van-Hai Bui\textsuperscript{1,*}}
\IEEEauthorblockA{\textsuperscript{1}Department of Electrical and Computer Engineering, University of Michigan-Dearborn, USA\\
\textsuperscript{2} School of Electrical and Electronic Engineering, Hanoi University of Industry, Hanoi, Vietnam.\\
\textsuperscript{3}Center for Environmental Intelligence, VinUniversity, Hanoi, Vietnam
}

\thanks{Corresponding author: Van-Hai Bui (vhbui@umich.edu).}}
\maketitle

\begin{abstract}
Oscillatory power components in AI data-center demand can produce large frequency deviations when they coincide with vulnerable grid modes. Mode-aware hybrid energy storage systems (HESSs) can selectively suppress these components, but their effectiveness degrades when the grid mode changes while the control target remains fixed. This paper proposes a closed-loop grid-mode-adaptive HESS control framework that integrates measurement-based modal identification with mode-selective control. A sliding-window N4SID estimator identifies the local grid dynamics from grid-side power and frequency measurements, and the resulting natural-frequency estimate is used to retune the modal suppression branch assigned to the supercapacitor. The framework is evaluated under a grid-mode shift from 0.50 to 1.35~Hz while maintaining equal physical damping and approximately equal load excitation at both frequencies. Compared with the fixed-target controller, the adaptive method reduces the post-change grid-side modal power component from 8.78 to 2.02~MW, corresponding to a 76.94\% reduction, and substantially lowers the resulting frequency deviation.
\end{abstract}

\section{Introduction}
As AI computing continues to grow, large data centers are becoming major electrical loads. Recent studies show that AI computing is an important driver of increasing data-center electricity demand~\cite{shehabi2024}. However, annual energy consumption and peak power demand alone do not fully describe their impact on power systems. AI training workloads also cause frequent changes in power demand, including rapid power ramps, checkpointing events, and short-term fluctuations. When many computing units change their activity at the same time, these power variations can create disturbances at the grid connection point~\cite{choukse2025power}. Therefore, it is important to understand how these disturbances affect grid dynamics.

To reduce load fluctuations, local energy storage can help control the power exchanged with the grid~\cite{mohammadi2026grid, liao2021adaptive}. In which, a battery energy storage system (BESS) can handle longer-term power variations, while a supercapacitor (SC) can respond quickly to short-term changes. By combining these two technologies, a hybrid energy storage system (HESS) can use the battery to handle slower power variations and the SC to handle faster fluctuations. Previous studies have explored battery–SC systems for data-center peak shaving and power capping~\cite{zheng2017}. More recently, coordinated HESS control has been proposed to reduce power ramps and fluctuations in data-center loads~\cite{ko2025, chen2026grid}.

However, reducing overall grid-side power fluctuations does not always reduce grid-frequency oscillations~\cite{chen2026grid}. The grid responds differently to power fluctuations at different frequencies and is more sensitive near its oscillatory modes. Therefore, even a small remaining power fluctuation near one of these modes can have a greater effect on grid frequency than a larger fluctuation at another frequency. As a result, two controllers with similar overall power-smoothing performance may have very different effects on frequency deviation. A control objective that focuses only on total power variation may therefore fail to reduce the fluctuations that have the greatest impact on grid frequency ~\cite{sarmadi2016}.

To address this problem, author in \cite{chen2026grid} proposed a mode-aware HESS control method that selectively suppresses power oscillations near the vulnerable grid frequency while retaining the conventional power-sharing strategy between the battery and SC. However, the modal target is treated as available to the controller and is not estimated online within the control loop. When grid operating conditions change, the oscillation mode may shift while the controller remains tuned to the previous target frequency, causing the SC to continue suppressing fluctuations at a frequency that is no longer critical. Therefore, the controller needs to update its target frequency based on the grid dynamics observed at the point of interconnection.

To address this issue, this paper proposes adaptive grid-mode-aware HESS control framework for AI data center oscillation mitigation. Specifically, grid-side power and frequency deviation are used as input–output measurements to identify the local grid dynamics. Based on these measurements, a low-order state-space model is estimated using N4SID~\cite{vanoverschee1994, dosiek2012mode}, from which the natural frequency and damping of the identified oscillation mode are obtained. Before updating the HESS modal filter, the estimated natural frequency is validated and smoothed. In this way, the HESS can maintain the power-sharing strategy between the battery and SC while adapting its control target to changes in the grid oscillation mode.

To the best of the authors' knowledge, the closed-loop integration of online grid-mode identification with mode-selective HESS control has not been investigated for AI data-center loads. This allows the controller to track changes in the grid oscillation mode without knowing its true frequency. To evaluate the proposed approach, three control cases are compared under the same load profile and grid damping conditions. This comparison helps show the benefits of tracking the grid mode while also highlighting a key limitation. Specifically, online identification requires sufficient measurement data after a mode change, which introduces a delay before the controller can adjust its target frequency. Therefore, the study evaluates not only the ability to reduce oscillations after the controller has adapted but also its performance during the transition period.

\section{System Configuration and Resonance Characteristics}
This section describes the AI data center--HESS--grid configuration, with emphasis on how grid-side power disturbances affect frequency deviation. It then discusses the resonance mechanism arising from the interaction between oscillatory data-center loads and the grid dynamics.

\subsection{Data Center--HESS--Grid Configuration}

\begin{figure}[!tb]
\centering
\includegraphics[scale=1.35]{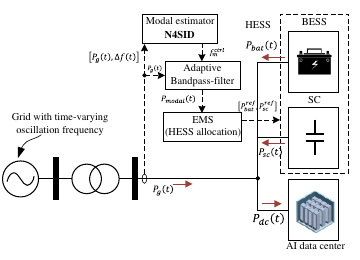}
\caption{AI data center--HESS--grid configuration with measurement-based grid-mode adaptation.}
\label{fig:config}
\end{figure}

The overall system and control configuration is shown in Fig.~\ref{fig:config} and consists of an AI data center, a supercapacitor (SC), a battery energy storage system (BESS), and an aggregate swing--governor grid model. The BESS and SC reshape the power seen by the grid according to
\begin{equation}
    P_g(t)=P_{dc}(t)-P_{bat}(t)-P_{sc}(t),
    \label{eq:balance}
\end{equation}
where $P_{dc}(t)$ denotes the data-center power demand, while positive $P_{bat}(t)$ and $P_{sc}(t)$ correspond to discharging power supplied by the storage devices.

Let
\begin{equation}
    \Delta P_g(t)=P_g(t)-\bar P_g
\end{equation}
denote the grid-side power disturbance around the operating point $\bar P_g$. The resulting grid-frequency dynamics are represented by the aggregate swing--governor model
\begin{align}
    M_g\Delta\dot f
    &=
    \Delta P_m
    -\frac{\Delta P_g}{P_{base}}
    -D_g\Delta f,
    \label{eq:swing}\\
    T_{gov}\Delta\dot P_m
    &=
    -\Delta P_m
    -\frac{\Delta f}{R_{gov}},
    \label{eq:governor}
\end{align}
where $M_g$ and $D_g$ denote the equivalent grid inertia and damping coefficients, respectively, $\Delta P_m$ is the mechanical-power deviation, $R_{gov}$ is the governor droop coefficient, $T_{gov}$ is the governor time constant, and $P_{base}$ is the power base.

For modal analysis, \eqref{eq:swing}--\eqref{eq:governor} can be compactly written as
\begin{equation}
    \dot{\mathbf{x}}
    =
    \mathbf{A}_g\mathbf{x}
    +
    \mathbf{B}_g u,
    \qquad
    y=\mathbf{C}_g\mathbf{x},
    \label{eq:grid_ss}
\end{equation}
where
\begin{equation}
    \mathbf{x}
    =
    \begin{bmatrix}
        \Delta f & \Delta P_m
    \end{bmatrix}^{\!T},
    \qquad
    u=\Delta P_g,
    \qquad
    y=\Delta f.
\end{equation}
Here, $\mathbf{A}_g$, $\mathbf{B}_g$, and $\mathbf{C}_g$ follow directly from
\eqref{eq:swing}--\eqref{eq:governor}.

The oscillatory characteristics of the grid are determined by the
eigenvalues of $\mathbf{A}_g$. For the second-order model above, the
natural frequency is
\begin{equation}
    \omega_n
    =
    \sqrt{
        \frac{D_g+1/R_{gov}}
        {M_gT_{gov}}
    },
    \qquad
    f_n=\frac{\omega_n}{2\pi}.
    \label{eq:natural_frequency}
\end{equation}

By varying the aggregate parameters $M_g$ and $D_g$, different grid operating conditions and corresponding oscillatory modes can be represented. When $\Delta P_g$ contains a sustained oscillatory component near a lightly damped natural frequency, the resulting frequency deviation can be strongly amplified. This resonance mechanism is examined in the next subsection.

\subsection{Resonance-Induced Frequency Amplification}

\begin{figure}[!tb]
\centering
\includegraphics[scale=0.8]{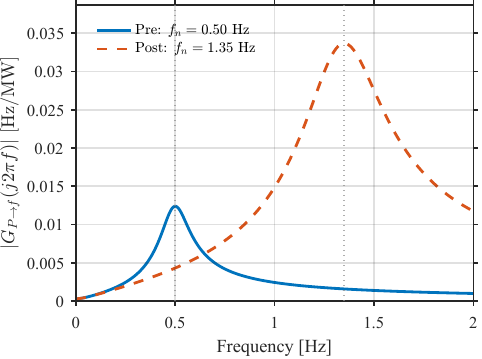}
\caption{Magnitude response from grid-side power disturbance to frequency deviation under the 0.50-Hz and 1.35-Hz grid operating conditions.}
\label{fig:magres}
\end{figure}

The grid impact of a power fluctuation depends not only on its
magnitude but also on its frequency. Eliminating $\Delta P_m$
from \eqref{eq:swing}--\eqref{eq:governor} gives the transfer
function from the grid-side power disturbance to frequency deviation:
\begin{equation}
\begin{gathered}
G_{P\rightarrow f}(s)
=
\frac{\Delta F(s)}{\Delta P_g(s)}
=
-\frac{T_{gov}s+1}
{P_{base}\mathcal{D}_g(s)},\\
\mathcal{D}_g(s)
=
M_gT_{gov}s^2
+
(M_g+D_gT_{gov})s
+
D_g+R_{gov}^{-1}.
\end{gathered}
\label{eq:transfer}
\end{equation}

For a sinusoidal grid-side power disturbance
$\Delta P_g(t)=A_p\sin(\omega_L t)$, the steady-state frequency
response has amplitude
\begin{equation}
    A_{\Delta f}
    =
    A_p
    \left|
    G_{P\rightarrow f}(j\omega_L)
    \right|.
    \label{eq:amplification}
\end{equation}
The corresponding frequency-response magnitude is
\begin{equation}
\left|G_{P\rightarrow f}(j\omega)\right|
=
\frac{\sqrt{1+(T_{gov}\omega)^2}}
{P_{base}\sqrt{a^2(\omega)+b^2(\omega)}} ,
\label{eq:frequency_gain}
\end{equation}
where $a(\omega)=D_g+R_{gov}^{-1}-M_gT_{gov}\omega^2$ and
$b(\omega)=(M_g+D_gT_{gov})\omega$.

For a lightly damped grid mode, \eqref{eq:frequency_gain} exhibits
a sensitivity peak near the modal frequency. Consequently, two
power oscillations with the same amplitude can produce substantially
different frequency deviations depending on their frequencies.
A component located near the modal peak produces a much larger
frequency response per unit of grid-side power than an off-modal
component. This behavior is illustrated in Fig.~\ref{fig:magres},
where the change in grid operating condition shifts the dominant
sensitivity from approximately $0.5$~Hz to $1.35$~Hz and also
changes the corresponding response magnitude.

This observation distinguishes broadband power smoothing from
modal oscillation mitigation. Reducing the overall variation of
$P_g$ does not necessarily remove the component to which the grid
is most dynamically sensitive. Therefore, effective mitigation
requires the HESS to selectively attenuate power oscillations near
the vulnerable grid mode. Because this mode can vary with the grid
operating condition, its target frequency must be updated online,
which motivates the adaptive framework developed in Section~III.

\section{Proposed Online Grid-Mode-Adaptive HESS Control Framework}
The proposed closed-loop grid-mode-adaptive HESS control architecture is illustrated in Fig.~\ref{fig:config}, where online modal estimation is integrated with the mode-selective HESS control branch.
\subsection{HESS Power Allocation and Modal Suppression}

The HESS controller consists of a broadband power-smoothing branch
and an additional mode-selective suppression branch. The former
reduces the overall data-center power variation, whereas the latter
specifically attenuates the grid-side power component near the
vulnerable oscillatory mode discussed in Section~II.

For the base smoothing action, a fixed fraction $\alpha_H$ of the
data-center power deviation from the operating level $P_c$ is assigned
to the HESS:
\begin{equation}
    P_{\mathrm{HESS}}^{base}
    =
    \alpha_H\left(P_{dc}-P_c\right),
    \qquad \alpha_H=0.75 .
    \label{eq:base_hess}
\end{equation}
Thus, $75\%$ of the load fluctuation around $P_c$ is assigned to the base HESS power command, while the remaining variation is allowed to appear at the
grid interface.

The resulting HESS power demand is then divided between the BESS
and SC according to their complementary dynamic characteristics.
A first-order low-pass filter
\begin{equation}
    F_b(s)
    =
    \frac{\omega_b}{s+\omega_b},
    \qquad
    \omega_b=2\pi f_b ,
    \label{eq:power_split}
\end{equation}
assigns the slower component to the BESS, while the complementary
higher-frequency component is supplied by the SC:
\begin{equation}
\begin{aligned}
    P_{bat}^{base} &= F_b(s)P_{\mathrm{HESS}}^{base},\\
    P_{sc}^{base}  &= \left[1-F_b(s)\right]P_{\mathrm{HESS}}^{base}.
\end{aligned}
\label{eq:base_allocation}
\end{equation}
This frequency-based allocation allows the BESS to accommodate
energy-dominant variations while reserving the faster SC response
for rapid load fluctuations.

However, the base smoothing branch reduces overall load variations without considering whether a particular frequency component is close to a vulnerable grid mode.
Motivated by the resonance characteristics discussed in
Section~II-B, an additional band-pass branch is therefore introduced
to extract the grid-side power component around the currently
targeted mode:
\begin{equation}
    W_m(s)
    =
    \frac{2\zeta_m\omega_m s}
    {s^2+2\zeta_m\omega_m s+\omega_m^2},
    \qquad
    \omega_m=2\pi f_m^{ctrl}.
    \label{eq:bpf}
\end{equation}
Since $W_m(0)=0$, the constant operating component of $P_g$ is
rejected by the band-pass filter, allowing the measured $P_g$ to be
used directly for modal extraction. Therefore, the extracted modal
component and its corresponding compensation command are
\begin{equation}
    z_m=W_m(s)P_g,
    \qquad
    P_{modal}=K_m z_m .
    \label{eq:modal_power}
\end{equation}

Because this compensation targets a relatively fast oscillatory
component, the modal command is assigned entirely to the SC branch.
The resulting storage power references are therefore
\begin{equation}
\begin{aligned}
    P_{bat}^{ref}
    &=
    P_{bat}^{base}+P_{bat}^{rec},\\
    P_{sc}^{ref}
    &=
    P_{sc}^{base}+P_{sc}^{rec}+P_{modal},
\end{aligned}
\label{eq:references}
\end{equation}
where $P_{bat}^{rec}$ and $P_{sc}^{rec}$ denote slow energy-recovery
terms that maintain the storage energy levels around their nominal
operating points. The resulting references are subject to the converter dynamics and
the power and energy constraints of each storage device, so the
actual storage powers may differ from their references.

Since the effectiveness of the modal branch depends on aligning
$f_m^{\mathrm{ctrl}}$ with the actual grid mode, the target frequency
must be updated from online measurements when the grid dynamics change.

\subsection{Online Modal Estimation Via N4SID Method}

To adapt the modal target to changing grid conditions, the current
grid mode is estimated directly from measurements at the point of
interconnection. At each update, a low-order model is identified
from recent grid-side power and frequency measurements over a
sliding window $\mathcal{W}$. The measured signals are centered as
\begin{equation}
    u_k=P_{g,k}-\overline{P_g}_{\mathcal W},
    \qquad
    y_k=\Delta f_k-\overline{\Delta f}_{\mathcal W},
    \label{eq:identdata}
\end{equation}
where $u_k$ represents the grid-side power disturbance and $y_k$
is the corresponding frequency deviation.

Using these input--output data, N4SID~\cite{vanoverschee1994}
identifies the discrete-time state-space model
\begin{equation}
    \hat{\mathbf{x}}_{k+1}
    =
    \hat{\mathbf{A}}_d\hat{\mathbf{x}}_k
    +
    \hat{\mathbf{B}}_d u_k,
    \qquad
    \hat y_k
    =
    \hat{\mathbf{C}}_d\hat{\mathbf{x}}_k
    +
    \hat{\mathbf{D}}_d u_k.
    \label{eq:ssid}
\end{equation}
This identified model provides a measurement-based discrete representation of the local grid dynamics in \eqref{eq:grid_ss}, without requiring the physical grid parameters to be known.
The oscillatory characteristics of the identified grid dynamics are
then obtained from the eigenvalues of $\hat{\mathbf{A}}_d$. For an
identified discrete pole $\lambda_d$, the corresponding continuous
pole is
\begin{equation}
    \lambda_c
    =
    \frac{\log(\lambda_d)}{T_s},
    \label{eq:continuous_pole}
\end{equation}
from which the natural frequency and damping ratio are calculated as
\begin{equation}
    \hat f_n
    =
    \frac{|\lambda_c|}{2\pi},
    \qquad
    \hat\zeta
    =
    -\frac{\Re(\lambda_c)}{|\lambda_c|}.
    \label{eq:modal_estimates}
\end{equation}

Physically admissible oscillatory poles are retained, and the
resulting natural-frequency estimate is conditioned before updating
the modal-filter center frequency $f_m^{\mathrm{ctrl}}$. This
conditioning prevents transient or noisy estimates from causing
unnecessary retuning, while preserving the ability of the controller
to follow sustained changes in the grid mode. The damping estimate
is used to support modal validation, whereas the modal-filter damping
$\zeta_m$ remains fixed.

Thus, the modal target is updated using only measured grid responses,
without requiring the true modal frequency or the underlying grid
parameters to be known by the controller.

\section{Numerical Results}

\subsection{Test Setup}

\begin{figure}[!tb]
\centering
\includegraphics[scale=0.85]{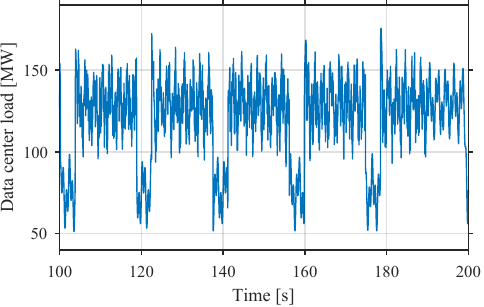}
\caption{Representative zoom-in of the synthetic AI data-center load profile generated based on \cite{choukse2025power}.}
\label{fig:load}
\end{figure}

The numerical study is designed to isolate the effect of grid-mode
variation on the performance of mode-aware HESS control. The
simulation lasts 300~s, and the grid natural frequency changes from
$0.50$ to $1.35$~Hz at $t=150$~s. The physical damping ratio is kept
at $\zeta_g=0.15$ in both operating conditions.

The same synthetic data-center demand, generated based on \cite{choukse2025power}, is applied in all cases. In addition to
the underlying workload variations, equal 12-MW oscillatory
components at $0.5$ and $1.35$~Hz are present throughout the
simulation. The zoom-in AI data center load profile from $100-200s$ is shown in Fig. \ref{fig:load}.

Three cases are considered:
\emph{No HESS}, in which the data-center load is directly supplied by
the grid;
\emph{Fixed}, in which the mode-selective HESS controller remains
tuned to $0.5$~Hz; and
\emph{Adaptive}, in which the modal target is updated from the online
N4SID estimate. Table~\ref{tab:parameters} summarizes the principal
parameters used in the study.

\begin{table}[!tb]
\caption{Principal System and Control Parameters}
\label{tab:parameters}
\centering
\footnotesize
\begin{tabular}{@{}ll@{}}
\toprule
Parameter & Value \\
\midrule
Sampling / simulation duration & 0.05 s / 300 s \\
Mode change time & 150 s \\
\hline
Aggregate grid $T_{gov}/R_{gov}/P_{base} $&   4s/3/10000MW\\
$[M_g,D_g]$ before / after &  $[0,0086,0,006]/[0,0012,0,0027]$ \\
$f_n$ before / after & 0.50 / 1.35 Hz \\
\hline
BESS rating / energy & 40 MW / 40 MWh \\
SC rating / energy & 20 MW / 0.30 MWh \\
$\alpha_H$; $P_c$; split cutoff $f_b$ & 0.75; 125 MW; 0.10 Hz \\
$K_m$; modal-filter damping $\zeta_m$ & 4; 0.10 \\
N4SID order / window / update & 2 / 80 s / 2 s \\
\bottomrule
\end{tabular}
\end{table}

\subsection{Online Mode Tracking and Grid-Frequency Mitigation}

\begin{figure}[!tb]
\centering
\includegraphics[width=0.88\columnwidth]{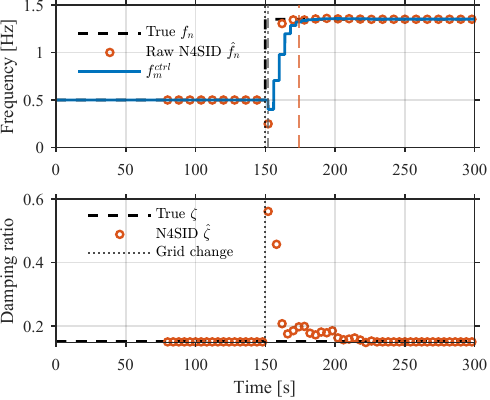}
\caption{Online grid-mode tracking during the transition from
$0.50$ to $1.35$~Hz. The estimated natural frequency is used to
retune the modal-filter center frequency.}
\label{fig:tracking}
\end{figure}

Figure~\ref{fig:tracking} shows the online modal estimates and the
resulting control target. Before the grid change, the identified
natural frequency closely follows the $0.5$-Hz mode. Immediately
after $t=150$~s, the sliding identification window contains data from
both operating conditions, producing a temporary deviation in the
estimated frequency and damping ratio. As post-change measurements
become dominant in the window, the estimated natural frequency
converges to the new $1.35$-Hz mode and the modal-filter center is
retuned accordingly.

\begin{figure}[!tb]
\centering
\includegraphics[width=0.90\columnwidth]{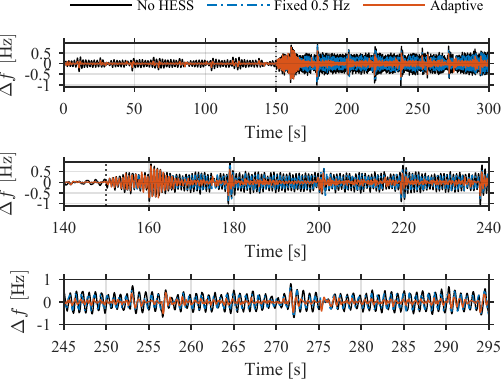}
\caption{Grid-frequency response for No HESS, Fixed, and Adaptive:
full simulation (top), mode-transition interval (middle), and
post-convergence interval (bottom).}
\label{fig:frequency}
\end{figure}

The effect of this retuning on the grid-frequency response is shown
in Fig.~\ref{fig:frequency}. Before the mode change, Fixed and
Adaptive methods exhibit essentially the same response because both are
centered at $0.5$~Hz. Once the grid mode moves to $1.35$~Hz, however,
the Fixed controller remains tuned to the former mode and loses its
targeted suppression capability. In contrast, Adaptive method follows the
identified mode and progressively restores oscillation mitigation.

Table~\ref{tab:results} quantifies the effect of modal retuning.
After the grid mode shifts to 1.35~Hz, the Fixed controller leaves
a 8.78-MW modal component in $P_g$, whereas Adaptive reduces it
to 2.02~MW, a 76.94\% reduction. The corresponding frequency
component decreases from 0.29 to 0.07~Hz, while the
post-change frequency RMS falls from 0.24 to 0.11~Hz.

The transition response also highlights a key limitation of the adaptive approach. The new grid mode must first be identified using a finite measurement window, resulting in an adaptation delay of approximately $20 s$ before the benefits of retuning become apparent.

\begin{table}[!tb]
\caption{Key Performance Comparison}
\label{tab:results}
\centering
\footnotesize
\begin{tabular}{@{}lccc@{}}
\toprule
Metric & No HESS & Fixed & Adaptive \\
\midrule
$A_{P_g}^{pre}$ (MW)       & 12.23  & 1.41   & 1.41 \\
$A_{P_g}^{post}$ (MW)      & 11.61  & 8.78  & 2.02 \\
$A_{\Delta f}^{post}$ (Hz) & 0.39 & 0.29 & 0.07 \\
$\Delta f_{\mathrm{RMS}}^{post}$ (Hz)
                           & 0.30 & 0.24 & 0.11 \\
\bottomrule
\end{tabular}
\end{table}

\subsection{Frequency-Domain Interpretation}

\begin{figure}[!tb]
\centering
\includegraphics[width=0.88\columnwidth]{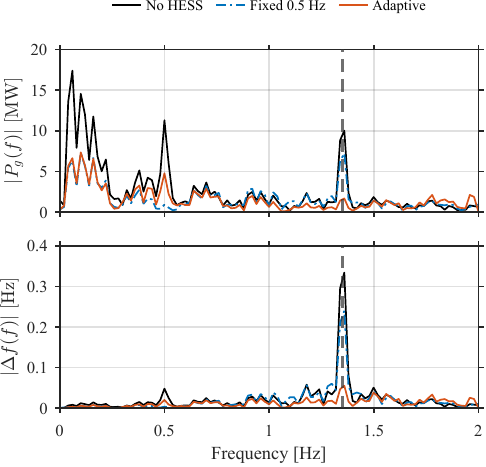}
\caption{Post-change amplitude spectra of grid-side power (top) and
frequency deviation (bottom). Fixed remains tuned to 0.5~Hz, whereas
Adaptive redirects modal suppression to the new 1.35-Hz grid mode.}
\label{fig:spectra}
\end{figure}

Figure~\ref{fig:spectra} explains the post-change response.
After the grid mode shifts to $1.35$~Hz, the Fixed controller
remains tuned to $0.5$~Hz and leaves a strong grid-side power
component near the new vulnerable mode. Adaptive instead retunes
the modal branch to $1.35$~Hz and substantially reduces both the
corresponding $P_g$ and $\Delta f$ components. This behavior is
consistent with equation \eqref{eq:amplification}, frequency impact depends
on both the disturbance magnitude and the grid sensitivity at that
frequency. The result therefore confirms that effective mitigation
requires targeting the current vulnerable mode rather than uniformly
smoothing the power spectrum.

\section{Conclusion}

This paper proposes a closed-loop grid-mode-adaptive HESS control
framework for mitigating oscillations induced by AI data-center
loads. The proposed approach combines measurement-based modal
identification with mode-selective HESS control, allowing the
targeted suppression frequency to follow changes in the grid
dynamics without requiring prior knowledge of the true grid mode.
When the grid natural frequency shifts from 0.50 to 1.35~Hz, the
fixed controller becomes mismatched, whereas the adaptive controller
retunes its modal branch and restores effective oscillation
mitigation. In the post-change regime, the adaptive method reduces
the targeted grid-side power component by 76.94\% relative to the
fixed controller and substantially lowers the resulting frequency
deviation. The main limitation is the adaptation delay introduced by
the finite identification window. Future work will focus on faster
modal tracking, multiple-mode and network-scale studies, and
real-time hardware-in-the-loop validation.

\section*{Acknowledgment}
This work was supported in part by the U.S. National Science Foundation (NSF) under Award 2552448 and in part  under Award 2509993.

\bibliographystyle{IEEEtran}
\bibliography{references}
\end{document}